\documentstyle[aps,prl,amssymb]{revtex}

\newcommand{\res}{\mathop{\rm res}\nolimits}
\newcommand{\Sl}{\mathop{\rm sl}\nolimits}

\begin{document}
\draft
\preprint{HEP/123-qed}
\title{Symmetries of the Kadomstev-Petviashvili Hierarchy}
\author{A. Yu. Orlov}
\address{Institute of Oceanology, Krasikova 23, Moscow 117218, Russia}
\author{P. Winternitz  \thanks{The research of P.W. was partially
supported by research grants from NSERC of Canada of FCAR du Qu\'ebec.
This investigation was conducted during A. Yu. O.'s visit to the CRM,
Universit\'e de Montr\'eal.}}
\address{Centre de Recherches Math\'ematiques, Universit\'e de
Montr\'eal, C.P. 6128 -- Succ. Centre-ville, Montr\'eal, Qu\'ebec H3C
3J7, Canada}
\maketitle
\begin{abstract}
The relation between the $\widehat{\Sl}(\infty)$ algebra of flows
commuting with the KP hierarchy and the Kac-Moody-Virasoro Lie point
symmetries of individual equations is established.  This is used to
calculate the point symmetries for all equations in the hierarchy.
\end{abstract}

\bigskip
The Kadomtsev-Petviashvili equation \cite{ref1} is of considerable
physical importance in the theory of wave propagation in such media as
shallow water, plasmas and others.  It is also the prototype of an
integrable equation \cite{ref2} in $2+1$ dimensions and as such, it
has a role to play in conformal quantum field theory and string
theories \cite{ref3}.  Like other integrable equations, the KP
equation can be viewed as a member of an infinite hierarchy of
mutually compatible integrable evolution equations, the KP
hierarchy\cite{ref4,ref5}.  Each equation in the hierarchy involves
the same space variables $x$ and $y$, but a different time variable
$t_N$ $(N \ge 3)$.

The purpose of this paper is to investigate the symmetries of the KP
hierarchy.  More specifically we will establish the relation between
the $\widehat\Sl(\infty)$ symmetries of the hierarchy of
equations\cite{ref4,ref5,ref6} and the Kac-Moody-Virasoro Lie point
symmetries of the KP equation itself \cite{ref7,ref8}.  In the process
we will show that each individual equation in the KP hierarchy has a
Kac-Moody-Virasoro Lie point symmetry algebra.  We shall present the
results without proofs, or details.  For those we refer to a
forthcoming article\cite{ref9}.  In order to present our results, we
shall make use of the language of pseudodifferential operators (PDO)
in one variable $f(x, \partial)$, with $\partial \equiv \partial
/\partial_x$, where $f$ is a function that can be expanded into a
Laurent series in $\partial$, including an arbitrary number of
positive and negative powers \cite{ref4,ref5,ref6}.  We shall use an
integral operator
\begin{equation}
\begin{array}{c}
\displaystyle K = 1 + \sum_{\ell = 1}^\infty
K_\ell(t_1,t_2,t_3,t_4,\ldots)\partial^{-\ell}\\
t_1 \equiv x, \quad t_2 \equiv y, \quad t_3 = t, \quad \partial \equiv
\partial_x.
\end{array}\label{eq1}
\end{equation}
The space of PDO is the direct sum of two subspaces $A = A_+ \dotplus
A_-$, where $A_+$ includes all nonnegative powers of $\partial$, $A_-$
all strictly negative ones.

The equations of the KP hierarchy can now be written in a compact form as
\begin{equation}
\frac{\partial K}{\partial t_n} = -( K \partial^n K^{-1} )_- K, \label{eq2}
\end{equation}
where the subscript minus means that only negative powers of
$\partial$ are retained.  Putting $\partial_{t_n} \equiv \partial_n$,
we can verify that all the flows (\ref{eq2}) commute: $[ \partial_n,
\partial_m]K = 0$.  For $n \in \Bbb Z^>$ fixed each operator equation
(\ref{eq2}) represents an infinite coupled set of equations for the
coefficients $K_\ell$ in the expansion (\ref{eq1}).  Various closed
subsets of equations can be obtained by fixing $n$ and $\ell$.  Thus,
taking $n \le 3$, and $\ell = 1,2,3$ and appropriately eliminating, we
get the potential KP equation for $w \equiv -2K_1$ (with $t_3 = t$)
\begin{equation}
w_{xt} = \frac{1}{4} w_{xxxx} + \frac{3}{2} w_x w_{xx} + \frac{3}{4}
w_{yy}. \label{eq3}
\end{equation}
Similarly, for $n=4$ we get the first higher KP equation
\begin{eqnarray}
\begin{array}{l}
w_{xxt_4} = \frac{1}{2} w_{xxxxy} + 3w_{xx} w_{xy} + 2w_x w_{xxy}\\
\displaystyle \qquad\qquad + w_{xxx} w_y + \frac{1}{2} w_{yyy}.
\end{array}
 \label{eq4}
\end{eqnarray}
The entire hierarchy can be written as a series of nonlinear and
nonlocal evolution equations on $w(x,y,t,t_4, \ldots)$, all involving
3 independent variables $x,y$ and $t_n$ (for $n \ge 3$)
\begin{equation}
\frac{\partial w}{\partial t_n} = 2 \res_\partial K \partial^n K^{-1}
\label{eq5}
\end{equation}
(where $\res_\partial$ means that we keep only the coefficient of
$\partial^{-1}$).

All equations of the KP hierarchy can be viewed as symmetries of the
KP equation.  They generate an abelian symmetry algebra and the
coefficients of the corresponding vector fields do not depend on the
independent variables.  Further symmetries exist that do not share the
above features: they commute with the KP flows, but not with each
other.  They can depend explicitly on $x,y$ and
$t_n$.\cite{ref6},..,\cite{ref10}  A detailed study of symmetries of
the KP hierarchy was performed in Ref.~6.  The corresponding equations
can be written as
\begin{equation}
\begin{array}{c}
[\partial_{m,n}, K] = -\bigl(K \hat x^n \partial^m K^{-1})_- K\\
\displaystyle \hat x = \sum_{k=1}^\infty k t_k \partial^{k-1}, \quad
n,m \in \Bbb Z
\end{array}\label{eq6}
\end{equation}
and they correspond to symmetries, since we have $[\partial_{mn},
\partial_k]K = 0$.  Their Lie algebra is isomorphic to $\widehat
\Sl(\infty)$, as can be seen by evaluating the commutator
$[\partial_{mn}, \partial_{m^\prime n^\prime}]K$.  The operator
$\partial_{mn}$ can be viewed as corresponding to differentiation with
respect to a new time.  The flow of $w$ with respect to this ``time''
is given by
\begin{equation}
\partial_{mn} w = 2 \res_\partial K \hat x^n \partial^m K^{-1}, \label{eq7}
\end{equation}
or equivalently
\begin{equation}
\partial_{mn} w = 2 \res_\lambda \lambda^m \frac{\partial^n
\varphi(\lambda)}{\partial  \lambda^n} \varphi^{\ast}(\lambda),
\label{eq8}
\end{equation}
where
\begin{equation}
\varphi(\lambda) = e^\zeta \Bigl( 1 + \sum_{k=1}^\infty \lambda^n K_n
\bigl(\vec t \bigr) \Bigr), \quad \zeta = \sum_{k=1}^\infty \lambda^k
t_k \label{eq9}
\end{equation}
is the formal Baker-Akhiezer function.\cite{ref5}

Lie point symmetries, at first glance, do not seem to fit into this
scheme.  Indeed, e.g.\ for the KP equation itself, the Lie point
symmetries correspond to a transformation $(x,y,t,w) \to (\tilde x,
\tilde y, \tilde t, \tilde w)$ such that $\tilde w(\tilde x, \tilde y,
\tilde t)$ is a solution, whenever $w(x,y,t)$ is one.  The Lie algebra
of this Lie group of local point transformations is represented by
vector fields of the form
\begin{equation}
\widehat V = \xi \partial_x + \eta \partial_y + \tau \partial_t + \phi
\partial_w, \label{eq10}
\end{equation}
where $\xi, \eta, \tau$ and $\phi$ are functions of $x,y,t$ and $w$.
The actual form of these functions is determined using a standard
algorithm,\cite{ref11} having nothing to do with the integrability of
the studied equation.  Once they are found, a ``symmetry'', in the
sense of a commuting flow (as in eq.~(\ref{eq5}), (\ref{eq7}), or
(\ref{eq8})) is obtained in the form
\begin{equation}
\frac{\partial w}{d \lambda} = \xi \frac{\partial w}{\partial x} +
\eta \frac{\partial w}{\partial y} + \tau \frac{\partial w}{\partial
t} - \phi. \label{eq11}
\end{equation}

Let us now show how the Lie point symmetries fit into the scheme fo
$\widetilde\Sl (\infty)$ symmetries (\ref{eq6}),...,(\ref{eq8}).  We
already know\cite{ref7} that for the KP equation the point symmetry
algebra involves arbitrary functions of time.  To include these we
replace the operators $c_{mn} = \hat x^n \partial^m$ by a functions
\begin{equation}
h_{\alpha, N} = \lambda^\alpha h \biggl( \frac{1}{N \lambda^{N-1}}
\frac{\partial}{\partial\lambda} \biggr) \label{eq12}
\end{equation}
where $\alpha$ and $N$ are integers and $h$ can be expanded into a
Laurent series in $\partial_E$, $E = \lambda^N$. Eq.~(\ref{eq8}) is
then replaced by
\begin{eqnarray}
\lefteqn{V^N(\alpha, h)w}\nonumber\\
&&\quad = 2 \res_\lambda \lambda^\alpha \biggl[ h\biggl( \frac{1}{N
\lambda^{N-1}} \frac{\partial}{\partial \lambda}
\biggr)\varphi(\lambda) \biggr]
\varphi^{\ast}(\lambda) \label{eq13}
\end{eqnarray}
where $V^N$ is a vector field, acting on $w$.  We shall use
eq.~(\ref{eq13}) to extract all {\em local} symmetries of the type
(\ref{eq13}), i.e. those not involving any integrals.  We expand the
function $h$ in eq.~(\ref{eq13}), as well as the Baker-Akhiezer
functions $\varphi(\lambda)$ and $\varphi^{\ast}(\lambda)$.  To avoid
nonlocal terms in the residue, we keep terms only upto or
$O(\lambda^{-N-2})$ in $\lambda$ (after pulling all derivatives to the
right).  The result of these operations is
\begin{eqnarray}
\begin{array}{l}
V^N(\alpha, h)w = \res \lambda^\alpha \biggl\{ h(t_N) + h^\prime(t_N)
\Bigl[ \frac{2}{N \lambda^{N-2}} y \\
\displaystyle  \quad + \frac{1}{N\lambda^{N-1}} x +
\frac{1}{2N\lambda^{N+1}} w \Bigr]\\
\displaystyle \quad
+ \frac{1}{2!} h^{\prime\prime}(t_N)  \Bigl[\frac{4}{N^2\lambda^{2N-4}} y^2 \\
\displaystyle \quad + \frac{4}{N^2 \lambda^{2N-3}} xy + \frac{1}{N^2
\lambda^{2N-2}} (x^2+4y-2Ny) \Bigr]\\
\displaystyle \quad + \frac{1}{3!} h^{\prime\prime\prime} (t_N) \Bigl[
\frac{8y^3}{N^3\lambda^{3N-6}}+ \frac{12 xy^2}{N^3 \lambda^{3N-5}} \Bigr]\\
\displaystyle \quad  + \frac{1}{4!} h^{\rm iv} (t_N) \frac{16y^4}{N^4
\lambda^{4N-8}} + O(\lambda^{-N-2}) \biggr\}\\
\displaystyle \quad \cdot \Bigl\{ 2 + \frac{w_x}{\lambda^2} +
\frac{w_y}{\lambda^3} + \ldots + \frac{w_{t_N}}{\lambda^{N+1}} +
O(\lambda^{-N-2}) \Bigr\}.
\end{array}\label{eq14}
\end{eqnarray}
Fixing $N \ge 3$ in eq.~(\ref{eq14}) corresponds to choosing an
equation in the KP hierarchy.  Again, $N=3$ is the KP equation itself.
The value of $\alpha$ corresponds to a specific symmetry.  Nonlocal
symmetries are avoided by making the restrictions on $\alpha$.  In
particular, for $\alpha = N$ we obtain a Virasoro algebra, for $-1 \le
\alpha \le 2$ we obtain a Kac-Moody type loop subalgebra of the
symmetry algebra (in both cases with no central extension).  Choosing
$N$ and $\alpha$ appropriately, we obtain the Lie point symmetries
from eq.~(\ref{eq14}) in the form of eq.~(\ref{eq11}), from which we
can read off the vector fields (\ref{eq10}) directly.  Let us run
through the individual cases and present the results in terms of
vector fields.
\begin{equation}
\begin{array}{l}
N = 3, \quad \alpha = 3,2,1,0,-1,\\
\displaystyle T(f) = f \partial_t + \frac{2}{3} yf^{\prime} \partial_y
+ \frac{1}{3} \biggl[ xf^\prime + \frac{2}{3} y^2 f^{\prime\prime}
\biggr] \partial_x,\\
\displaystyle  \quad  -\frac{1}{3} \biggl[ wf^\prime + \frac{1}{3} x^2
f^{\prime\prime} + \frac{4}{3^2} xy^2 f^{\prime\prime\prime}
+\frac{4}{3^4} y^4 f^{\rm iv} \biggr] \partial_w,\\
\displaystyle Y(g) = g \partial_y + \frac{2}{3} yg^\prime \partial_x -
\frac{4}{9} y \biggl( xg^{\prime\prime} + \frac{2}{9} y^2
g^{\prime\prime\prime}\biggr)\partial_w,\\
\displaystyle X(h) = h \partial_x - \frac{2}{3} \biggl( xh^\prime +
\frac{2}{3} y^2 h^{\prime\prime} \biggr) \partial_w,\\
W(k) = ky \partial_w,\quad U(\ell) = \ell \partial_w.
\end{array} \label{eq15}
\end{equation}

\begin{equation}
\begin{array}{l}
N = 4, \quad \alpha = 4,2,1,-1\\
\displaystyle T(f) = f \partial_t + \frac{1}{4} f^\prime (x \partial_x
+ 2y \partial_y) \\
\displaystyle \qquad\qquad\qquad - \frac{1}{4} (wf^\prime + xy
f^{\prime\prime}) \partial_w,\\
\displaystyle Y(g) = g \partial_y - \frac{1}{2} xg^\prime \partial_w,\\
\displaystyle X(h) = h \partial_x - y h^\prime \partial_w, \quad W(k)
= k \partial_w.
\end{array} \label{eq16}
\end{equation}

\begin{equation}
\begin{array}{l}
N= 5, \quad \alpha =5,2,1,-1,\\
\displaystyle T(f) = f \partial_t + \frac{1}{5} f^\prime (x\partial_x
+ 2y \partial_y) \\
\displaystyle \qquad\qquad\qquad - \biggl( \frac{1}{5} wf^\prime +
\frac{4}{25} y^2 f^{\prime\prime} \biggr) \partial_w,\\
\displaystyle Y(g) = g \partial_y - \frac{4}{5} g^\prime y \partial_w,
\quad X(h) = h \partial_x, \\
\displaystyle W(k) = k\partial_w,
\end{array} \label{eq17}
\end{equation}

\begin{equation}
\begin{array}{l}
N \ge 6, \quad \alpha =6,2,1,-1,\\
\displaystyle T(f) = f \partial_t + \frac{1}{N} f^\prime (x \partial_x
+ 2y \partial_y - w \partial_w),\\
\displaystyle Y(g) = g \partial_y, \quad X(h) = h \partial_x,\\
\displaystyle W(k) = k \partial_w.
\end{array} \label{eq18}
\end{equation}
In eq.~(\ref{eq15}),\dots,(\ref{eq18}) all of the functions $f,g,h,k$
and $\ell$ depend on the corresponding time $t_N$ and are otherwise
arbitrary.

Eq.~(\ref{eq15}) for the potential KP agrees with ref.~\cite{ref9},
the cases $N \ge 4$ are new.  The first field $T(f)$ in each case
corresponds to the Virasoro algebra, the remaining ones to loop
algebras, based on solvable Lie algebras.  Notice that the Lie point
symmetries go far beyond the obvious ``physical'' symmetries.  Indeed
time translations, and dilations are obtained from the Virasoro
algebra $T(f)$ as $T(1)$ and $T(t)$, respectively.  In general, $T(f)$
generates arbitrary reparametrizations of time.  Similarly, $X(1)$ is
an $x$-translation, $X(t)$ a Galilei transformation; $Y(1)$ a
$y$-translation, $Y(t)$ a Galilei transformation for $N \ge 4$.  For
$N = 3$ it is a Galilei transformation combined with a rotation.
Finally, $W(k)$ and $U(\ell)$ generate gauge transformations.

We mention that all known integrable equations in $2+1$ dimensions
have Kac-Moody-Virasoro Lie point symmetry algebras (Davey-Stwartson
equation, 3-wave equations, two dimensional Toda lattice,
etc.).\cite{ref12}  The results presented here show that this is no
coincidence and give credence to the conjecture that the occurrence of
Kac-Moody-Virasoro symmetries can be used as an integrability
indicator for PDEs in 3-dimensions of space-time.

\end{document}